# Geometrothermodynamics of a gravitating system with axially symmetric metric

**Grushevskaya H.V., Krylova N.G.**

Belarusian State University, Minsk, Belarus

*E-mail: grushevskaja@bsu.by, krylovang@bsu.by;*

Theory of the first-order phase transition in contact statistical manifold is proposed to describe interface systems. The theory has not limitations of known Van der Waals-like phase transition theories. Based on this approach, geometrothermodynamics of gravitating systems with axially symmetric metrics is investigated. It has been shown that such geometrothermodynamics occurs in space-time with Newman–Unti–Tamburino-like metric.

## 1. Introduction

To date, there are known important experimental facts in cosmology: a nonlinear dependence of galaxies recession on magnitude of the red shift, a nonlinear dependence of supernova luminosity on the red shift, gravitational lensing of galaxies which does not associated with fluctuations, the existence of voids in galaxies distribution [1]. These facts can be explained through the existence of space-time universe regions which is expanding at an accelerating rate. Gravitating matter in such regions does not interact (or interacts weakly) with ordinary substance and is known as dark matter. Beside the fact that our universe is in the phase of accelerated expansion, the universe is a flat one, and therefore vacuum energy cosmological constant $\Lambda$ has to be nonzero: $\Lambda \neq 0$ [2-5]. $\Lambda$-term within Einstein equations describes the dark energy (see, for example, [6] and references therein] which can be interpreted as a fluid with negative pressure $p$, $p < 0$ [7]. The pressure takes on a negative value at the first-order phase transition while the matter is in a metastable state. Because of this, it has been proposed in [8-9] to use the thermodynamics at gravity modeling. $\Lambda$CDM (Lambda cold dark matter) cosmological models [10] and electrically charged AdS (anti de Sitter) black holes [11-12] exhibit a liquid-gas like first order (1st-order) phase transition culminating in critical points that resemble the phase diagram of a Van der Waals fluid [13].

However, the Van der Waals–Maxwell gas model is not invariant with respect to the physically meaningful Legendre transformations [14-15]. Moreover, Einstein equations are not conformally invariant because of Weyl tensor does not vanish. The thermodynamics of cosmological models, such as Friedmann–Lemaître–Robertson–Walker ones, includes effects of the non-null Weyl tensor (anisotropic pressure) by introducing a viscosity parameter and relaxation time (inverse expansion coefficient) [16-17]. But, relaxation processes are absent in the Van der Waals–Maxwell gas model. Thus, a more realistic theory of the 1st-order phase transition has to be used to develop geometrothermodynamic approaches to cosmology.

In [18] we have proposed a theory of the 1st-order phase transition in a contact statistical manifold that describes interface systems with an electrocapillary mechanism of the energy dissipation and with relaxation time distributions; and this theory has not limitations which are in the Van der Waals-like phase transition theories. The triad $\{\vec{r}, \dot{\vec{r}}, \vec{\xi}\}$ for space-time ($r$, $t$) defines our contact manifold which is called the thermodynamic phase space $\mathcal{T} = \{\vec{r}, \dot{\vec{r}}, \vec{\xi}\}$, $\dot{\vec{r}} = \vec{r}_\xi \left(\dfrac{ds}{d\xi}\right)^{-1}$ [19-21]. The tangent space $T_{\mathcal{T}} = \{\dot{\vec{r}}, \vec{\xi}\}$ is a subspace of the bundle $\mathcal{T}$. The tangent space $T_{\mathcal{T}}$ has the vertical (hidden) $\mathcal{V}_\xi$ and horizontal ($\mathcal{H}$) direct sum components: $T_{\mathcal{T}} = \mathcal{V}_\xi \oplus \mathcal{H}$. Thus, $\mathcal{V}_\xi$ is the vertical subspace generated by $\vec{\xi}$, and $\mathcal{H}$ is the horizontal distribution. The bundle $\mathcal{T}$ of our problem can be redefined by the dynamical-system phase-space $\{\vec{r}, \vec{r}_\xi\}$, which is augmented by the subspace spanned by the tangent vector $\dfrac{ds}{d\xi}$:

$\mathcal{T} = \left\{\vec{r}, \vec{r}_\xi, \dfrac{ds}{d\xi}\right\}$. The bundle $\mathcal{T}$ includes the four-dimensional (4D) space of the two pairs $\{\vec{r}, \vec{r}_\xi\}$ and admits a non-trivial connection determined by the vector $\dfrac{\partial}{\partial s}$. Here $\dfrac{\partial}{\partial s}$ is a Reeb vector field of the contact manifold (see [22-24] and references therein). Therefore, $\mathcal{T}$ is a fifth-dimension (5D) contact manifold. Because the 1$^{\text{st}}$-order phase transition proceeds on the interface, the metric is an axially symmetric one.

In the paper, utilizing approach considered above we study the geometrothermodynamics of gravitating systems with an axially symmetric metric which is similar to Newman–Unti–Tamburino (NUT) one.

## 2. A geometro-thermodynamical background of problem

An entropy production $\Delta S$ in the first-order phase transition proceeding on the interface [18] reads

$$\Delta S = \sum_a \Delta S_a = -\sum_a F_a \Delta t_a + \sum_a E_a \Delta t_a, \tag{1}$$

where $\Delta S_a$ is the entropy production for the $a$-th process,

$$-\sum_{a=1}^N F_a \Delta t_a = \Delta s \sum_{a=1}^N \tau_a F_a = \Delta s F = \ln Z, \tag{2}$$

$$\sum_{a=1}^N E_a \Delta t_a = -\Delta s \sum_{a=1}^N \tau_a E_a = -\sum_{a=1}^N \dfrac{\partial \ln Z}{\partial \Delta t_a} \Delta t_a = -N \Delta s \dfrac{\partial \ln Z}{\partial \Delta s}. \tag{3}$$

$Z$ is the statistical sum

$$Z = \int \prod_{a=1}^N \mathcal{D} X_a\, p(\vec{r}_1,...,\vec{r}_N,t) \exp\left\{\Delta s \sum_a \tau_a \left[H(X_a) + i\Gamma(X_a)\right]\right\},$$

$\int \prod_{a=1}^N \mathcal{D} X_a$ represents the path integral $\int \mathcal{D} X = \int \prod_{a=1}^N \mathcal{D} X_a$ over configurations $X_a$, $X_a = \{\vec{r}_1(t_1);...;\vec{r}_{a-1}(t_{a-1}); \vec{r}_a(t_a + \Delta t_a); \vec{r}_{a+1}(t_{a+1});...;\vec{r}_N(t_N)\}$. The Hamiltonian $H(X_a)$ and the decay rate $\Gamma(X_a)$ of each individual relaxation process are renormalized by its relaxation time $\tau_a$, the role of inverse temperature $\beta$ in the transition state would have been played by the parameter $(-\Delta s)$. The formulas (2) and (3) are determined the Helmholtz free energy $F$ and the internal energy $E$ for the first-order phase transition as

$$F = \sum_{a=1}^N \tau_a F_a = \dfrac{1}{\Delta s} \ln Z, \quad E = \sum_{a=1}^N \tau_a E_a / N = \dfrac{\partial \ln Z}{\partial \Delta s}. \tag{4}$$

Expanding the expression (1) in series over small parameter $\Delta s$ as

$$-\sum_a F_a \Delta t_a = \ln Z \big|_{\Delta t_a = 0} + \sum_a \Delta t_a \left[\dfrac{\int p\{X_a\}(H(X_a) + i\Gamma(X_a)) DX_a}{Z}\right]_{\Delta t_a = 0} = \sum_a \Delta t_a\, C, \tag{5}$$

$$\sum_a E_a \Delta t_a = -\sum_a \dfrac{\partial \ln Z}{\partial \Delta t_a} \Delta t_a = -\dfrac{1}{Z} \sum_a \Delta s \int p\{X_a\}[\tau_a(H(X_a) + i\Gamma(X_a))] DX_a; \tag{6}$$

and passing to the limit $\dot{\xi} = \frac{\partial t}{\partial s} = \lim_{\Delta s \to 0} \frac{\Delta t}{\Delta s}$ one can find an action on the statistical manifold of the 1st-order phase transition:

$$\Delta S = \int \left[ \dot{\xi} C - L(\dot{\xi}) \right] ds.$$

In a case of electrocapillary loss, the Lagrangian $L(\dot{\xi}) \equiv L(\vec{r}, \dot{\vec{r}}, \dot{\xi})$ reads in polar coordinates $(r, \varphi)$

$$L(\vec{r}, \dot{\vec{r}}, \dot{\xi}) = -p|V|r^5 e^{\frac{2|V|\xi}{r}} \frac{\dot{\xi}^2}{\dot{r}} + U(\xi, r)\dot{\xi} + m\frac{(\dot{r}^2 + r^2\dot{\varphi}^2)}{2\dot{\xi}}, \tag{7}$$

and, respectively, the action $dl$ on the statistical manifold is written in the following form:

$$dl = L(\vec{r}, \dot{\vec{r}}, \dot{\xi}) ds = \left( -p|V|r^5 e^{\frac{2|V|\xi}{r}} \frac{\dot{\xi}^2}{\dot{r}} + U(\xi, r)\dot{\xi} + m\frac{(\dot{r}^2 + r^2\dot{\varphi}^2)}{2\dot{\xi}} \right) ds, \tag{8}$$

where $\dot{r} = \frac{dr}{ds}$ and $\dot{\varphi} = \frac{d\varphi}{ds}$,

$$U(\xi, r) = p \left\{ \left[ -\frac{4}{3}r^5 + \frac{16}{15}(|V|\xi)r^4 + \frac{1}{30}(|V|\xi)^2 r^3 + \frac{1}{45}(|V|\xi)^3 r^2 \right. \right.$$
$$\left. \left. + \frac{1}{45}(|V|\xi)^4 r + \frac{2}{45}(|V|\xi)^5 \right] e^{\frac{2|V|\xi}{r}} - \frac{4}{45} \frac{(|V|\xi)^6}{r} \mathrm{Ei}\left( \frac{2|V|\xi}{r} \right) \right\},$$

$|V|$ is a modulus of compression (or expansion) rate, $p$ and $m$ are model parameters.

## 3. Generalized NUT-metric

The NUT metric reads

$$dl^2 = \left( 1 - \frac{r_g r + 2n^2}{r^2 + n^2} \right) \left( dx^0 + 4n\sin\frac{\theta}{2} d\varphi \right)^2 - \frac{dr^2}{1 - \frac{r_g r + 2n^2}{r^2 + n^2}} - (r^2 + n^2)(d\theta^2 + \sin^2\theta d\varphi^2), \tag{9}$$

where $n$ is a NUT parameter, $r_g$ is a gravitational radius, $x_0$ is a time coordinate; $r$, $\theta$, $\varphi$ are spatial spherical coordinates. In the case of $r^2, r_g r \gg n^2$ this metric behaves like Schwarzschild metric or Reissner–Nordström one if a term $\frac{q^2 + g^2}{r^2 + n^2}$ is added. Here $q$ and $g$ are electric and magnetic charges, respectively. Since a following approximation $\Phi^{-1}(r)r^2 \approx (r^2 + n^2)$, $\Phi(r) = 1 - \frac{r_g r + 2n^2}{r^2 + n^2}$ holds, one can transform the expression (9) as

$$dl^2 = \Phi(r)\left( dx^0 + 4n\sin^2(\theta/2) d\varphi \right)^2 - \frac{dr^2}{\Phi(r)} - \frac{r^2}{\Phi(r)}(d\theta^2 + \sin^2\theta d\varphi^2). \tag{10}$$

At $r^2, r_g r < n^2$ one can add a correction $(\Lambda/3)(r^2 + 5n^2) - \frac{8(\Lambda/3)n^4 - q^2 - g^2}{r^2 + n^2}$ to the metric (10). Then the NUT-metric transforms to the form:

$$dl^2 = \tilde{\Phi}(r)\left( dx^0 + 4n\sin^2(\theta/2) d\varphi \right)^2 - \frac{dr^2}{\tilde{\Phi}(r)} - \frac{r^2}{\tilde{\Phi}(r)}(d\theta^2 + \sin^2\theta d\varphi^2), \tag{11}$$

where $\tilde{\Phi}(r) = 1 - \dfrac{r_g r + 2n^2}{r^2 + n^2} + \dfrac{\Lambda}{3}(r^2 + 5n^2) - \dfrac{8(\Lambda/3)n^4 - q^2 - g^2}{r^2 + n^2}$, $\Lambda$ is a cosmological constant.

While $r^2, r_g r \ll n^2$, $r \sim r_g$, the substitution $r^2/\tilde{\Phi}(r) \approx (r^2 + n^2)/\left(1 + \dfrac{5\Lambda}{3}n^4\right) = R^2$, $dr = dR$ leads to the metric (11) obtains the features of anti de Sitter metric.

## 4. An axially symmetric metric of thermodynamic phase space entering a 5-dimensional contact manifold

Based on the metric function (8) one can construct the following metric:

$$dl^2 = L(\vec{r}, \vec{\dot{r}}, \dot{\xi})ds^2 = -p|V|r^5 e^{\frac{2|V|\xi}{r}}\dfrac{\dot{\xi}}{\dot{r}}d\xi\, ds + U(\xi, r)\dot{\xi}\, ds^2 + m\dfrac{(\dot{r}^2 + r^2\dot{\varphi}^2)}{2\dot{\xi}}ds^2. \qquad (12)$$

A space-time metric can be obtained from the metric (12) by the Wick rotation $\xi = -i\tau$ and $s = -i x^0$:

$$d\tilde{l}^2 = p|V|r^5 \exp\left(\dfrac{-i2|V|\tau}{r}\right)\dfrac{\dot{\xi}}{\dot{r}}d\tau\, dx^0 - U(-i\tau, r)\dot{\xi}(dx^0)^2 - \dfrac{m}{2\dot{\xi}}(dr^2 + r^2 d\varphi^2), \qquad (13)$$

where $dr = \dot{r} dx^0$, $d\varphi = \dot{\varphi} dx^0$. Let $\dot{\xi}$ be a known function $F(r)$: $\dot{\xi} = F(r)m/2$. Then, the metric (13) can be considered as a metric of 4D-surface $\theta = \pi/2$ in the contact 5D-manifold:

$$d\tilde{l}^2 = \dfrac{m}{2}F(r)\left[-U(-i\tau, r)(dx^0)^2 + \exp\left(\dfrac{-i2|V|\tau}{r}\right)\dfrac{p|V|r^5}{\dot{r}}d\tau\, dx^0\right] - \dfrac{\left[dr^2 + r^2\left(d\theta^2 + \sin^2\theta d\varphi^2\right)\right]}{F(r)}, \quad(14)$$

$\theta = \pi/2$.

Let us choose some another 4D-surface in the contact 5D-manifold in a such way to the angle $\theta$ takes arbitrary values, while ranges of variables $r$, $\varphi$ are dynamically limited by the variable $\tau$. This can be realized in the following way. Let us impose a gauge condition on the scalar field $\phi \equiv e^{\frac{-i|V|\tau}{r}}$:

$$\dfrac{4p}{45}\dfrac{(|V|\tau)^6}{r} = 1. \qquad (15)$$

We consider the case of small values of rate $|V| \to 0$ when the function $\mathrm{Ei}\left(-i\dfrac{2|V|x^0}{r}\right)$ turns to a constant value $\mathrm{Re}\,\mathrm{Ei}(i0) = -\infty$, $\mathrm{Im}\,\mathrm{Ei}(i0) = \pi/2$. Then, while the condition (15) is imposed, the scalar field $\phi$ oscillates strongly,

$$U(-i\tau, r) = -\mathrm{Ei}(i0) \text{ at } |V| \to 0, \text{ and } r = \dfrac{4p}{45}(|V|\tau)^6. \qquad (16)$$

Due to the constraint (15), one has $\dot{r} = \dfrac{dr}{d\tau}\dfrac{d\tau}{-i dx^0} = \dfrac{24p}{45}|V|^6 \tau^5 \dfrac{d\tau}{-i dx^0}$. Then changing $(dx^0)^2 \to (m/2)(dx^0)^2$ the metric (14) takes the form

$$dl^2_{renorm} = F(r)\left[\mathrm{Im}\,\mathrm{Ei}(i0) - i\exp\left(\dfrac{-i2|V|\tau}{r}\right)\dfrac{45}{24}\dfrac{r^5}{|V|^5 \tau^5}\right](dx^0)^2 - \dfrac{\left[dr^2 + r^2\left(d\theta^2 + \sin^2\theta d\varphi^2\right)\right]}{F(r)}, (17)$$

where $dl^2_{renorm} = d\tilde{l}^2 - \mathrm{Re}\,\mathrm{Ei}(i0) F(r)(dx^0)^2$.

Now, one has to exclude the 5-th coordinate $\tau$. A phase $\frac{2|V|\tau}{r}$ in exponential term $\exp\left(\frac{-i2|V|\tau}{r}\right)$ of 5-th dimension varies in the range from 0 up to $\pi$. Therefore, the phase $\frac{|V|\tau}{r}$ is defined by the angle $\theta/2$:

$$dl_{renorm}^2 = F(r)\left\{\text{Im Ei}(i0) - i\left(\cos\frac{\theta}{2} - i\sin\frac{\theta}{2}\right)^2 \frac{45}{24}\frac{r^5}{|V|^5 \tau^5}\right\}(dx^0)^2 - \frac{\left[dr^2 + r^2\left(d\theta^2 + \sin^2\theta d\varphi^2\right)\right]}{F(r)}.$$
(18)

At $\tau \to \infty$ the expression (18) is an expansion over a small parameter $1/\tau$:

$$dl_{renorm}^2 = F(r)\left[d\tilde{x}^0 - i\frac{\left(\cos\frac{\theta}{2} - i\sin\frac{\theta}{2}\right)^2}{48\,\text{Im Ei}(i0)}\frac{45 r^5 d\tilde{x}^0}{|V|^5 \tau^5}\right]^2 - \frac{\left[dr^2 + r^2\left(d\theta^2 + \sin^2\theta d\varphi^2\right)\right]}{F(r)} \quad (19)$$

where $d\tilde{x}^0 = (\text{Im Ei})^{1/2}(i0)dx^0$. Then, taking into account that $0 \leq \frac{r}{|V|\tau} \leq 1$ to impose the constraint of the form $2\pi\left(\frac{r}{|V|\tau}\right)^5 d\tilde{x}^0 = d\varphi$ on the angle $\varphi$, one gets

$$dl_{renorm}^2 = F(r)\left[d\tilde{x}^0 - \frac{45i}{24\pi}\left(\cot\frac{\theta}{2} - i\right)^2 \sin^2\frac{\theta}{2} d\varphi/(2\pi)\right]^2 - \frac{\left[dr^2 + r^2\left(d\theta^2 + \sin^2\theta d\varphi^2\right)\right]}{F(r)}. \quad (20)$$

Comparison between (11) and (20) gives the expression for the NUT-parameter, which becomes an imaginary constant at $\theta \to \pi, 0$:

$$n\sin^2\frac{\theta}{2}\bigg|_{\theta \to \pi,0} \equiv -\frac{45i}{192\pi^2}\sin^2\frac{\theta}{2}\left(\cot\frac{\theta}{2} - i\right)^2\bigg|_{\theta \to \pi,0} = \frac{45i}{192\pi^2}.$$

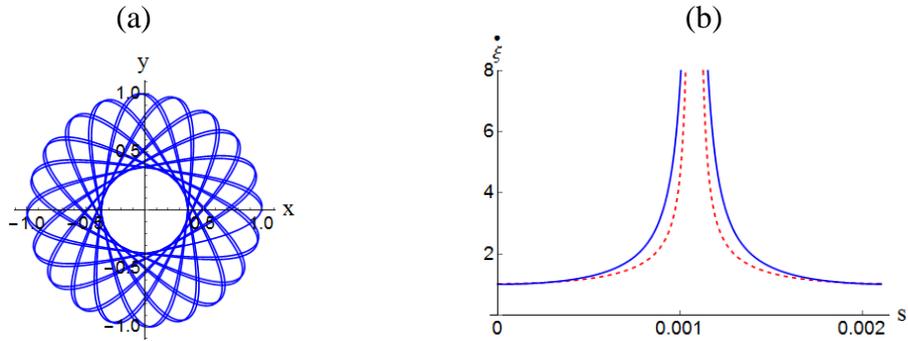

Fig. 1. (a) Geodesic; (b) the dependence of $\dot{\xi}(s)$ (blue solid curve) and its approximation by $\dot{\xi}(s) = -0.8 + \frac{1}{r(s) - 0.45}$ (red dashed curve).

Geodesic in the contact manifold with the metric (12) and an estimation of dependence of $\dot{\xi}$ at $|V| \to 0$ are shown in fig. 1. The simulation results and the analytic calculations performed above prove that such Schwarzschild-like geodesics behave similarly to that of the NUT metric.

## 5. Discussion and conclusion

So, using of the contact statistical 5-dimensional manifold, which describes phase transitions proceeding in the interface systems with an electrocapillary mechanism of energy dissipation, allows us to construct the theory of cosmological models with axially symmetric metrics. These metrics are that of 4-dimensional surfaces in the 5D manifold. We has shown that the NUT-theory parameter *n* is the gauge parameter of the scalar field which plays a role of fifth dimension. The metrics in 5D world are similar to the (anti) de Sitter metrics at large values of *n*.